# LINEAR OPTICAL SUB-DOPPLER RAMSEY RESONANCES IN ULTRATHIN GAS CELLS

## AZAD Ch. IZMAILOV

*Institute of Physics, Azerbaijan Republic Ministry of Sciences and Education,
Baku, AZ-1073, Azerbaijan
e-mail: izmailov57@yahoo.com*

The paper theoretically establishes and studies the sub-Doppler linear optical Ramsey resonances that arise under certain conditions near centers of optical transitions in the absorption of a sufficiently weak monochromatic light beam during its stationary propagation in the normal direction through an ultrathin gas cell whose internal thickness is less than or of the order of the wavelength of this radiation. We consider a situation when the cross section of this beam consists of two coaxial spatially separated regions, with the absorption signal being detected in a comparatively narrow central part of the beam. The paper studies the significant dependence of the linear optical Ramsey resonances, that arise in such an absorption spectrum, on the distance between these regions, as well as on the phase difference between them. In particular, it is shown that if this phase difference is close to π, then instead of absorption, amplification of the central part of the incident beam can occur. The Ramsey resonances under consideration are most clearly manifested when the internal thickness of the gas cell is equal to a small half-integer number of wavelengths of the resonant radiation. However, these resonances do not arise if this thickness is equal to an integer number of such waves. The established Ramsey resonances, under certain conditions, can find application in ultrahigh resolution atomic (molecular) spectroscopy, as well as effective references in compact optical frequency standards.

## 1. INTRODUCTION

In ultrahigh resolution atomic (molecular) spectroscopy, a considerable reduction of the time-of-flight broadening of spectral lines could be achieved by the realization of the Ramsey method of separated electromagnetic fields [1,2]. This method can be used to obtain spectral absorption resonances without Doppler broadening with a width inverse to the time of flight of atoms in a collimated beam between spatially separated electromagnetic fields. However, to manifest such ultra-narrow Ramsey resonances in the linear optical regime, the divergence of such a beam must be less than or on the order of the ratio $(\lambda/D)$ of the wavelength $\lambda$ of optical radiation to the distance $D$ between the light fields. For example, for characteristic values of $\lambda \sim 1 \mu m$ and $D \sim 10 cm$, such a small divergence $\lambda/D \sim 10^{-5}$ of an atomic beam is very difficult to realize.

At the same time, ultrathin gas cells with a characteristic transverse size $D \sim 1 cm$ and an internal thickness $L$ down to tens of nanometers have been manufactured [3]. Spectroscopy of such microscopic (nanoscopic) gas cells gives access to a new regime of gas dynamics in which the thermodynamical equilibrium becomes cell-specific. The corresponding atom-light interaction exhibits peculiar features, including a fully transient and coherent response, a strongly anisotropic atom time-of-flight and an explored space smaller than the wavelength. Transmission spectroscopy of ultrathin gas cells allows one to study the field of Quantum Electrodynamics inside a dielectric nanocavity (where light is confined in a sub-wavelength space), as well as nanophysics and nanotechnology with "quasi interaction-free" atoms [4]. Moreover, ultrathin cells, with their opportunity for Doppler free, linear spectroscopy in a tightly confined environment, are potential candidates for relatively compact, portable optical clocks.

Indeed, in such cells with rarefied vapors of alkali metal atoms, sub-Doppler resonances caused by the Dicke effect have been recorded in the linear optical mode [5-7]. Earlier, during a theoretical study of given resonances in these thin cells, it was assumed that the incident light wave was plane [5-9]. However, with spatial separation of coherent light fields in the ultrathin gas cells under consideration, it is possible also to create the necessary conditions for the manifestation of linear optical Ramsey resonances in the absorption of the incident radiation. Indeed, a set of coherently excited atoms in such a cell is an analogue of a compact atomic beam with a characteristic divergence $(L/D)$ [9,10], which can be significantly less than the limit $(\lambda/D)$ required for the occurrence of Ramsey resonances.

Therefore, in this paper, we theoretically investigate the features of linear optical absorption in a plane ultrathin cylindrical gas cell for a monochromatic light beam incident in the normal direction. It is assumed that the cross section of this beam consists of two coaxial spatially separated regions, according to the scheme in Fig. 1. A beam with such a spatial configuration can be obtained in practice from a conventional monochromatic laser beam by means of known light beam shaping applications [11]. It is considered that the detection of radiation absorption is carried out in a relatively narrow central part of this beam with the radius $r_1 \ll r_2$ (Fig.1). It is assumed also that the frequency of the incident light beam is close to the central frequency of the optical transition $a \rightarrow b$ that couples the ground atomic quantum state $a$ with the sufficiently long-lived (metastable) excited level $b$. Then, under certain conditions, sub-Doppler Ramsey resonances arise in such absorption, caused directly by

131, H.Javid ave, AZ-1073, Baku
Institute of Physics
E-mail: jophphysics@gmail.com

the transfer of optically induced atomic polarization from the peripheral region of this beam to the central detection region (Fig.1).

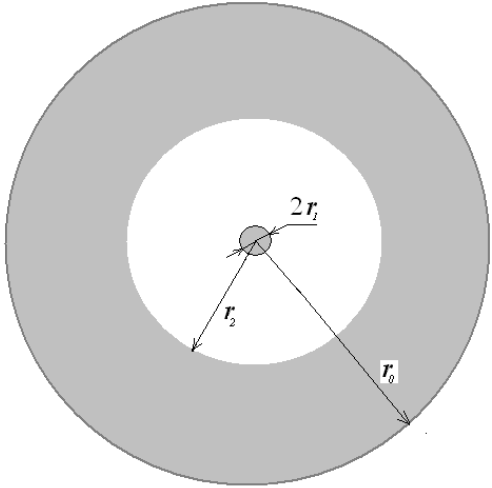

*Figure 1.* Cross-sectional scheme of an ultra-thin cylindrical gas cell with the radius $r_0$, where coaxial regions of the monochromatic light beam incident in the normal direction are marked in gray. It is believed that absorption detection is carried out in the narrow central region of this beam with the radius $r_1 \ll r_2$.

In the following Section 2, within the framework of the semiclassical theory of interaction of coherent radiation with rarefied gas atoms, relations for the linear optical absorption of a traveling monochromatic light beam in an ultrathin cylindrical gas cell are obtained. Based on these relations, in Section 3, the sub-Doppler Ramsey resonances arising in such absorption are established and investigated. In particular, the significant dependence of these resonances on the spatial separation and phase difference of the coaxial regions of the incident light beam (Fig.1), as well as on the internal thickness of the gas cell, is analyzed. In conclusion (Section 4), the main findings of this work are presented.

## 2. BASIC RELATIONSHIPS

Let us consider an atomic (or molecular) gas that is contained in a transparent cylindrical cell with fairly small internal thickness $L$ and radius $r_0 \gg L$ (Fig.1). The gas is assumed to be sufficiently rarefied that interatomic collisions can be neglected. Suppose that a weak monochromatic laser beam is propagating in the normal direction through the gas medium (along the $z$ axis) under steady state conditions. Its electric field is

$$\mathcal{E}(\mathbf{R}, t) = \mathbf{e} * E(r) * \exp[i(\omega t - kz)] + c.c., \quad (1)$$

where $\mathbf{R} = (x, y, z)$ is the three-dimensional radius vector, ω and e are the radiation frequency and unit vector of polarization, respectively, $k = \omega/c$ is the wave number, $E(r)$ is the radiation strength that has cylindrical transversal distribution on the distance $r = \sqrt{x^2 + y^2}$ from the beam axis. The frequency ω of the beam (1) is close to the central frequency $\omega_0$ of the electric-dipole transition $a \to b$ that couples the ground atomic quantum state $a$ with the sufficiently long-lived (metastable) excited level $b$, both of which are assumed to be nondegenerate. Then the interaction of the weak wave (1) with atoms of the rarefied gas in the linear optical regime can be described by the well-known equation for the induced optical coherence $\rho_{ab}$ between quantum states $a$ and $b$ with populations $n_a$ and $n_b$ [12]:

$$\frac{\partial \rho_{ab}}{\partial t} + \mathbf{v} \frac{\partial \rho_{ab}}{\partial \mathbf{R}} + (\gamma - i\omega_0)\rho_{ab} = i \frac{(\mathcal{E}\mathbf{d})}{\hbar} (n_b - n_a) F(\mathbf{v}), \quad (2)$$

where $\mathbf{v}$ and $\mathbf{R}$ are the velocity and coordinate of an atomic particle, respectively, $\mathbf{d}$ is the dipole moment matrix element for the resonant transition $a \to b$ with the natural halfwidth $\gamma$ of the spectral line, $F(\mathbf{v})$ is the Maxwell distribution on the atomic velocity v. Substituting the expression

$$\rho_{ab} = \xi_{ab} * \exp[i(\omega t - kz)] \quad (3)$$

into the relation (2) taking into account the formula (1), we obtain the following equation for the value $\xi_{ab}$:

$$\mathbf{v} \frac{\partial \xi_{ab}}{\partial \mathbf{R}} + [\gamma + i(\delta - kv_z)]\xi_{ab} = iE(r) \frac{(\mathbf{e}\mathbf{d})}{\hbar} (n_b - n_a) F(\mathbf{v}), \quad (4)$$

where $\delta = (\omega - \omega_0)$ is the frequency detuning of the wave (1) from the resonance, $v_z$ is the atomic velocity projection on the wave vector. It is assumed that the induced linear optical coherence of atoms disappears when they collide with the cell walls. Then, from Eq.(4), we obtain the following solution for the component of this coherence $\xi_{ab}$ at the central axis of the cylindrical cell (Fig.1) with coordinates $r = 0$ and $0 \leq z \leq L$:

$$\xi_{ab} = i \frac{(\mathbf{e}\mathbf{d})}{\hbar v_t} (n_b - n_a) * F_l(v_z) * F_t(v_t) \{\eta(v_z) * \eta(r_0 v_z - zv_t) * p(zv_t/v_z) + [\eta(v_z) * \eta(zv_t - r_0 v_z) + \eta(-v_z) * \eta(zv_t - Lv_t - r_0 v_z)] * p(r_0) + \eta(-v_z)\eta(r_0 v_z - zv_t + Lv_t) * p[(z - L)v_t/v_z]\}, \quad (5)$$

where $p(h) = \int_0^h E(r) \exp(-\Lambda r / v_t) dr$, $\Lambda = \gamma + i(\delta - kv)$, $\eta(g)$ is the step function ($\eta(g) = 1$ for $g \geq 0$ and $\eta(g) = 0$ if $g < 0$), $F_l(v_z)$ and $F_t(v_t)$ are Maxwell distributions for the longitudinal $v_z$ and transversal $v_t$ components of the atomic velocity, respectively:

$$F_l(v_z) = \pi^{-0.5} \cdot u^{-1} \cdot \exp(-v_z^2 \cdot u^{-2}), \quad F_t(v_t) = 2v_t \cdot u^{-2} \cdot \exp(-v_t^2 \cdot u^{-2}), \quad (6)$$

$u$ is the most probable atomic speed in the gas. Next, in accordance with the scheme in Fig.1, we will carry out calculations of the optical coherence $\xi_{ab}$ for the following distribution of the electric field over the cross-section of the incident light beam (1):

$$E(r) = E_1 \cdot \eta(r_1 - r) + E_2 \cdot \exp(i\varphi) \cdot \eta(r - r_2) \cdot \eta(r_0 - r), \quad (7)$$

where the relation $r_0 > r_2 \gg r_1$ takes place, $E_1$ and $E_2$ are constant values of the field strength in the central and peripheral regions of the beam, respectively, and $\varphi$ is a constant phase difference between these light regions. Radiation of such a spatial configuration (7) can be obtained in practice from a conventional monochromatic laser beam by means of known laser beam shaping applications [11]. For the linear optical processes under consideration with a field strength (7), this coherence $\xi_{ab}$ (3), (4) is the sum of two terms:

$$\xi_{ab} = E_1 \sigma_{ab}^{(1)} + E_2 \sigma_{ab}^{(2)}, \quad (8)$$

where values $\sigma_{ab}^{(1)}$ and $\sigma_{ab}^{(2)}$ do not depend on the radiation intensity. According to (8), influence of the peripheral region of the light beam (Fig.1) on its absorption in the narrow coaxial central part can be determined experimentally directly by the detected difference of such absorption signals at a given strength $E_2$ and when $E_2 = 0$. Then the spectrum of such a difference signal will be determined only by the value $\sigma_{ab}^{(2)}$ in (8). Since the detection of beam absorption is carried out in its fairly narrow central part with a radius $r_1 \ll r_2$ (Fig.1), then in the calculations of this value $\sigma_{ab}^{(2)}$ (8), it is possible to use the expression for the optical coherence $\xi_{ab}$ (5), calculated for the central axis of the gas cell.

In our case, the variation of the intensity $I_1 = E_1^2$ of the detected central part of the light beam (1), (7) is described by the known relationship [12]:

$$\frac{dI_1}{dz} = \left(\frac{4\pi\omega}{c}\right) E_1 * \text{Im}(\mathbf{Pe}), \quad (9)$$

where $\mathbf{Pe}$ is the projection of the $\mathbf{P}$ component of the gas polarization on the vector $\mathbf{e}$ (1) which is determined by the optical coherence $\xi_{ab}$ (8):

$$\mathbf{P} = \mathbf{d} \int \xi_{ab} \, d\mathbf{v}. \quad (10)$$

Based on relations (7)-(10), we obtain the following expression for the absorbed radiation power in the central region of the light beam (1), (7) with the cross-sectional area $\pi r_1^2$ (Fig.1):

$$J = J_1 + J_2, \quad (11)$$

where

$$J_1 = \left(\frac{4\pi\omega}{c}\right) \pi r_1^2 (\mathbf{ed}) \left\{ \int_0^L \left[\int \text{Im}(\sigma_{ab}^{(1)}) dv\right] dz \right\} E_1^2, \quad (12)$$

$$J_2 = \left(\frac{4\pi\omega}{c}\right) \pi r_1^2 (\mathbf{ed}) \left\{ \int_0^L \left[\int \text{Im}(\sigma_{ab}^{(2)}) dv\right] dz \right\} E_1 E_2. \quad (13)$$

The above-mentioned method of detecting the difference absorption signal can be used to experimentally determine the contribution to the total power $J$ (11) directly from the term $J_2 \sim E_1 E_2$ (13). This term is caused by the transfer to the center of the beam (1), (7) of the atomic polarization, induced in the peripheral region of the light beam (Fig.1). The contribution of the term $J_2$ (13) to the detected absorption signal $J$ (11) will be most significant at the relation $E_2 \gg E_1$ (7) and at a sufficiently small radius $r_1$ of the central detectable region of the light beam compared to the size $(r_0 - r_2)$ of its peripheral region (Fig.1). Further, precisely at such conditions, we will investigate the spectral structures associated with this term $J_2$ (13), where Ramsey resonances arise under certain conditions.

## 3. DISCUSSION OF RESULTS

Our analysis will be performed using, as example, the data for the intercombination transition $^1S_0 - ^3P_1$ (at the wavelength $\lambda = 689.5$ nm) between the ground term $^1S_0$ and the metastable level $^3P_1$ of strontium atoms [13], whose most probable speed in the gas at a temperature about $400^0$C is $u \approx 350$ m/s. A rather large lifetime $\tau \approx 2 \cdot 10^{-5}$s of the excited level $^3P_1$ determines the comparatively small natural halfwidth $\gamma = 0.5/\tau \approx 25$kHz of the spectral line of the $^1S_0 - ^3P_1$ transition.

Fig.2 shows the spectral dependences of $J_2(\delta)$ (13) for three different values of the phase difference $\varphi = 0, \pi$ and $0.5\pi$ between the coaxial regions of the incident light beam (1), (7). For $\varphi = 0$, the high-contrast sub-Doppler peak appears with the center $\delta = 0$ in the $J_2(\delta)$ dependence (curve 1 in Fig. 2). This Ramsey resonance is caused directly by the increase of the induced polarization of the gas medium in the detected part of the light beam due to the entry of optical coherence of atoms from the in-phase peripheral region of this beam (Fig.1). Note that characteristic oscillations arise in the $J_2(\delta)$ dependence in the vicinity of such a narrow Ramsey resonance. Curve 2 in Fig.2 was calculated for the case of the phase difference $\varphi = \pi$ (7), when the coaxial regions of the light beam (1) are in antiphase (Fig.1). Then the narrow dip corresponding to the amplification of the radiation appears in the center of the $J_2(\delta)$ dependence. In the case of an intermediate phase difference $\varphi = 0.5\pi$, the sub-Doppler resonance of the dispersion form appears in the vicinity of the frequency detuning $\delta = 0$ of the $J_2(\delta)$ dependence (curve 3 in Fig. 2).

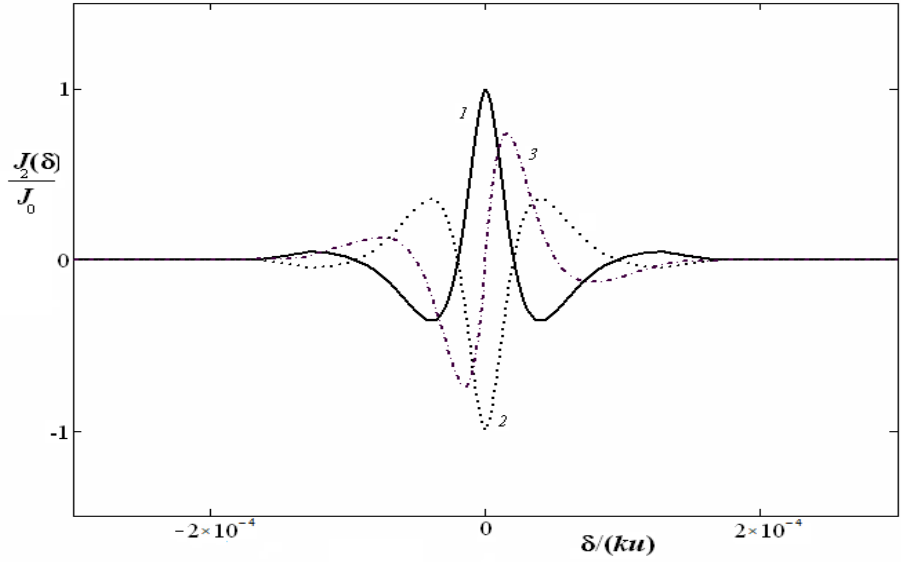

*Figure 2.* Dependence of the value $J_2(\delta)$ on the frequency detuning $\delta$ (in fractions of the Doppler broadening $ku$) for the light beam consisting of two coaxial regions (Fig.1) with the phase difference of $\varphi = 0$ (curve 1), $\pi$ (2) and $\varphi = 0.5\pi$ (3) when $L = 0.1\mu m$, $r_2 = 5mm$ and $r_0 = 15mm$. The value of $J_2(\delta)$ is normalized to its value $J_0$ at $\delta = 0$ and $\varphi = 0$.

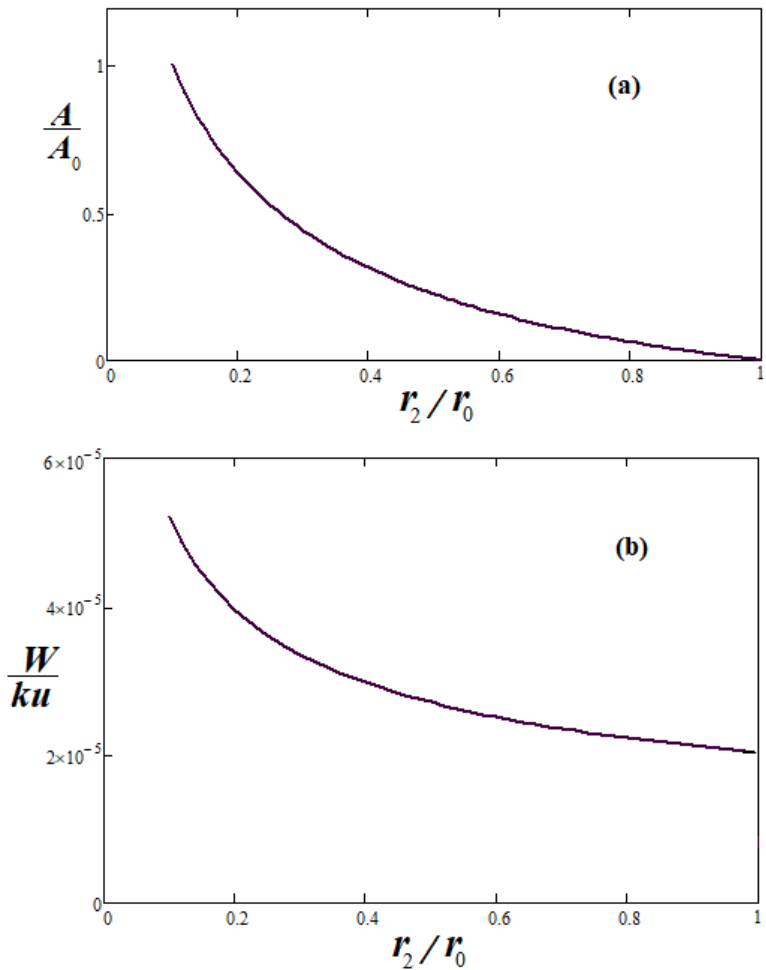

*Figure 3.* Dependence of the amplitude $A$ (a) and width $W$ (b) (in fractions of the Doppler broadening $ku$) of the Ramsey resonance, described by the function $J_2(\delta)$ (13), on the radius $r_2$ (Fig.1), when $\varphi = 0$, $L = 0.1\mu m$ and $r_0 = 10mm$. The value of $A$ is normalized to its value of $A_0$ at $r_2 = 0.1r_0$.

In Fig.3, for the phase difference $\varphi = 0$ (7), the dependences of the amplitude $A$ of the Ramsey resonance described by the quantity $J_2$ (13) at $\delta = 0$ and its effective width $W$ (at the half-maximum) on the radius $r_2$ (Fig.1) (7) are presented. Then, it is obvious that the amplitude $A$ decreases with increasing of $r_2$ down to 0 at $r_2 \rightarrow r_0$ (Fig. 3a). At the same time, the removal of the peripheral region of the light beam from its coaxial central part (Fig. 1) will lead to an increase in the flight time of optically excited atoms between these regions. Thus, due to a decrease in the time-of-flight broadening of the spectral line, a narrowing of the Ramsey resonance occurs (Fig. 3b). Note that the minimum width of the Ramsey resonance $W \approx 2 \cdot 10^{-5} ku \approx$ 64 kHz at $r_2 \rightarrow r_0$ in Fig. 3b is close to its natural width $2\gamma \approx$ 50 kHz for the parameters of the intercombination transition $^1S_0 - {^3P_1}$ of the strontium atom used in our calculations. Note also that in the case of the phase difference $\varphi = \pi$ (7), there are dependences of the characteristics of the Ramsey resonance similar to Fig. 3, which, however, has the opposite sign in absorption compared to the case of $\varphi = 0$ (Fig. 2).

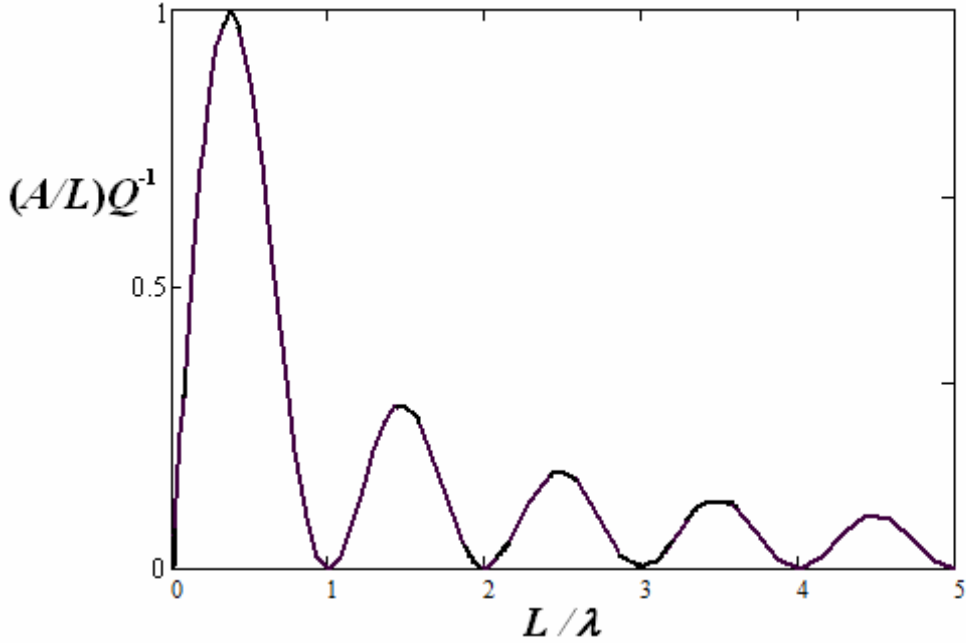

*Figure 4.* Dependence of the value of *A/L* on the internal thickness of the gas cell *L* (in units of the wavelength $\lambda$), when $\varphi = 0, r_2 = $ 1mm and $r_0$=10mm. The value of *A/L* is normalized to its maximum value $Q$ at $L = 0.5\lambda$.

Fig. 4 shows the dependences on the internal thickness $L$ of the gas cell for the ratio *A/L* of the amplitude *A* of the Ramsey resonance to the value $L$ for certain parameters of the light beam (7) with the phase difference $\varphi = 0$. Given dependences have the character of damped oscillations. In this case, the value *A/L* reaches maximum values when the thickness $L$ is equal to a half-integer number of wavelengths $\lambda$ of the resonant radiation and *A*=0 if $L$ is equal to an integer number of $\lambda$. Note that similar oscillatory dependences on the thickness $L$ are also characteristic for linear optical Dicke resonances detected in ultrathin gas cells by means of an ordinary monochromatic light wave with a plane front [5-9]. These oscillations are caused by the addition of induced optical coherences for the entire set of atoms along the cell. For an internal thickness $L$ equal to an integer number of wavelengths $\lambda$, the contribution of such coherences, shifted in phase by $\pi$, will be the same in magnitude but opposite in sign. As a result, the influence of its spatially separated peripheral region will not be manifested in the detected linear optical absorption of the central part of the light beam (Fig.1). At the same time, if the internal thickness $L$ of the gas cell is equal to a half-integer number $\lambda$, then the in-phase contribution of such light-induced coherences of atoms will reach a maximum and the Ramsey resonances will manifest themselves most clearly. However, the Ramsey resonances under consideration do not arise if $L \gg \lambda$. Note that in the case of the phase difference $\varphi = \pi$ (7), there are dependences similar to Fig. 4.

Despite the fact that figures 2-4 were obtained on the basis of numerical calculations for the characteristic parameters of the intercombination optical transition of $^1S_0 - {^3P_1}$ of strontium atoms, the same qualitative results were obtained by author in calculations for similar intercombination transitions of Mg, Ca and Ba atoms [13].

**4. CONCLUSION**

We have established and analyzed linear optical Ramsey resonances that arise under certain conditions near centers of optical transitions in the absorption of a sufficiently weak monochromatic light beam during its stationary propagation in the normal direction through an ultrathin cell with rarefied gas. These sub-Doppler Ramsey resonances are caused directly by the transfer

15

of optically induced atomic polarization from the peripheral region of this beam to its central detection region (Fig.1). The considered Ramsey resonances are directly related to the term $J_2 \sim E_1 E_2$ (13) in the total detected absorption power $J = J_1 + J_2$ (11). This term $J_2$ may be dominant in the absorption signal when the internal thickness of the gas cell $L$ is close to a small half-integer number of wavelengths $\lambda$ of the incident radiation (for example, if $L/\lambda \approx 0.5, 1.5, 2.5$), and also at a sufficiently small radius $r_1$ of the central detectable region of the light beam compared to the size $(r_0 - r_2)$ of its peripheral region (Fig.1). In addition, for the intensities of these coaxial regions, the relation $E_2 \gg E_1$ (7) should be fulfilled. Structure of studied Ramsey resonances essentially depends also on the phase difference $\varphi$ (7) between these light regions. In particular, at the phase difference $\varphi = \pi$, instead of absorption, amplification of the central detectable part of the incident beam can occur due to the considered Ramsey resonances.

The narrow sub-Doppler linear optical Ramsey resonances established and analyzed in this work can find application in atomic (molecular) ultrahigh resolution spectroscopy [4]. In addition, these resonances can be effective reference for high-precision compact frequency standards [14] due to the negligible dependence of the light shift and broadening of these resonances on the intensity of the sufficiently weak laser radiation inducing them.

___________________________________